\documentclass[aps,pre,twocolumn,superscriptaddress,floatfix,showpacs]{revtex4}
\usepackage{graphicx}
\usepackage{subfigure}

\bibstyle{apsrev.bib}

\begin{document}

\title{Collective roughening of elastic lines with hard core interaction\\
in a disordered environment} 

\author{Viljo Pet\"aj\"a}
\affiliation{
Helsinki University of Techn., Lab. of Physics, 
P.O.Box 1100, 02015 HUT, Finland}
\author{Deok-Sun Lee}
\affiliation{
Theoretische Physik, Universit\"at des Saarlandes, 
66041 Saarbr\"ucken, Germany}
\author{Mikko Alava}
\affiliation{
Helsinki University of Techn., Lab. of Physics, 
P.O.Box 1100, 02015 HUT, Finland}
\author{Heiko Rieger}
\affiliation{
Theoretische Physik, Universit\"at des Saarlandes, 
66041 Saarbr\"ucken, Germany}

\begin{abstract}
We investigate by exact optimization methods the roughening
of two and three-dimensional systems of elastic lines with point
disorder and hard-core repulsion with open boundary conditions.  
In 2d we find logarithmic behavior whereas in 3d simple random walk
 -like behavior. The line 'forests' become
asymptotically completely {\it entangled} as the system height is
increased at fixed line density due to increasing line wandering.
\end{abstract}

\pacs{74.60.Ge, 05.40.-a, 74.62.Dh} 

\maketitle

\newcommand{\bc}{\begin{center}}
\newcommand{\ec}{\end{center}}
\newcommand{\be}{\begin{equation}}
\newcommand{\ee}{\end{equation}}
\newcommand{\ba}{\begin{array}}
\newcommand{\ea}{\end{array}}
\newcommand{\beqn}{\begin{eqnarray}}
\newcommand{\eeqn}{\end{eqnarray}}

\section{Introduction}

The paradigmatic model for the mixed or Shubnikov phase of a
super-conductor in a magnetic field is a system of interacting elastic
lines, representing the magnetic vortex lines threading the
superconductor along the field direction \cite{blatter}.  The number
of lines per unit area perpendicular to the field axis, also called
the line density $\rho$, is given by the magnetic field strength and
also defines the typical line-to-line distance. The interaction
between the lines is repulsive and long ranged but exponentially cut
off beyond a screening length, $\lambda$.  As long as this screening
length is larger than the typical line distance $a=1/\rho^{1/d}$
(i.e. $\lambda>a$) at low temperatures the lines arrange in a periodic
pattern, the Abrikosov flux line lattice. In the presence of quenched
disorder such as it is produced by crystal lattice defects or
impurities this long range order is lost and for weak disorder or low
magnetic field strengths transformed into quasi-long-range order, the
Bragg glass phase
\cite{bragg-glass}. For strong disorder the arrangement of lines is
highly irregular and the system is in a state that is called the
vortex glass \cite{vortex-glass}.

In this paper we want to study the case when the screening length
$\lambda$ is much {\it smaller} than the typical line distance
$\lambda\ll a$). This corresponds to high magnetic field or strong
disorder and small values of $\lambda$. From a theoretical point this
limiting case is particularly interesting since it corresponds to a
situation in which the system cannot be described by an elastic theory
any more, which is based on the assumption of weak disorder and low
line density. Perturbative approaches that start from a ground state
with long range order and assume small fluctuations turn out to be
inappropriate in this strongly disordered situation and one has to
rely on numerical calculations to study the low temperature
properties.

Here we study the extreme limit of a system of lines in a disordered
environment with hard core repulsion. Although it is inspired by
strongly disordered superconductors in a magnetic field it describes
the more general situation of a dense system of one-dimensional
(line-like) objects that compete for the energetically most favorite
locations in the matrix they live in but repel each other via a hard
core exclusion principle - for instance long polymers stretching from
one side of a very inhomogeneous sample to the opposite side.  We use
a simplified (directed) polymer model and study the disorder averaged
ground state properties of this system. To calculate the exact ground
states we use an optimization algorithm, the
minimum-cost-flow-algorithm, which determines the exact optimal
solution in polynomial time \cite{opt-review}. 

We focus on the roughness properties of the lines (i.e the typical
transverse fluctuations of the lines on their way from the top to the
bottom of the system) and will show that in two dimensions the steric
repulsion between the lines is sufficient to produce the same scaling
behavior of the roughness as predicted by the elastic theories for
long range interactions, weak disorder and low line-densities. In
particular in 2d the hard core repulsion leads to collective
rearrangements of the lines that yield a roughness that increases
logarithmically with system size, also called super-rough behavior
\cite{superrough,elastic2d}. 
In three space dimensions the situation is totally
different: Elastic theories predict that the roughness increases with
the square-root of the logarithm of the system size \cite{elastic3d},
whereas we find for the case we
consider that the lines become more or less transparent for each other
and can wander transversally from one side of the sample to the other.
Steric repulsion alone is thus not sufficient to restrict the
transverse fluctuations of a line system in 3d.

The paper is organized as follows: In the next section we will
introduce the model and specify the methods by which we compute and
analyze its ground states. In section 3 and 4 we will present our
results for the roughness of the 2d and 3d system, respectively, and
present the corresponding finite size scaling forms. In section 5 we
consider various generalization of the disorder, in particular weak
and anisotropic randomness, and derive the scaling laws that govern
the crossover behavior to the universal scaling forms reported in the
previous sections. Section 6 contains a discussion and an outlook.

\section{Model}

We consider the following model of a system of interacting lines in a
two- or three-dimensional disordered environment: The lines live on the
bonds of a simple cubic lattice with a lateral width $L$ and a
longitudinal height $H$ (i.e.\ $M=L \times H$ or $M=L \times L \times
H$ lattices sites in 2d or 3d, respectively) with free boundary
conditions in all directions. The lines start at the bottom plane and
end at the top plane, if necessary the entrance and exit points can be
fixed. The number $N$ of lines threading the sample is fixed by a
prescribed density $\rho=N/L$ and $\rho=N/L^2$ in two and three
dimensions respectively. We restrict ourselves to hard-core
interactions between the lines, which means that their configuration
is specified by bond-variables $n_i\in\{0,1\}$, $n_i=1$ indicating
that a line segment occupies a bond with index $i$, and $n_i=0$
indicating that no line segment occupies this bond.

We model the disordered environment by assigning a random (potential)
energy $e_i\in[0,1]$ (uniformly distributed) to each bond $i$, such
that the total energy of the line configuration is given by
\be
{\mathcal H}=\sum_i e_i n_i\;.
\label{ham}
\ee
In section 5 the distribution of $e_i$ is modified from the presented
above in order to extend our analysis also to the cases of weak and
anisotropic disorder.

In a continuum limit the system of interacting elastic lines is
described by the following Hamiltonian:
\beqn
{\mathcal H} = {\displaystyle\sum_{i=1}^N}
{\displaystyle \int_0^H dz }
\Bigl\{ \frac{\gamma}{2}\left[\frac{d{\bf r}_i}{dz}\right]^2
+V_r[{\bf r}_i(z),z] \nonumber \\
+\sum_{j(\ne i)}V_{\rm int}[{\bf r}_i(z)-{\bf r}_j(z)]
\Bigr\}\; ,
\label{cont}
\eeqn
where ${\bf r}_i(z)$ denotes the transversal coordinate at
longitudinal height $z$ of the $i$-th flux line.  Our numerical model
corresponds to the case where the interactions $V_{\rm int}[{\bf
r}_i(z)-{\bf r}_j(z)]$ are hard core repulsive and the
$\delta$-correlated disorder potential $V_r[{\bf r}_i(z),z]$ has to be
strong compared to the elastic energy that is proportional to
$\gamma$.

At low temperature the line configurations will be dominated by the
disorder and thermal fluctuations are negligible. Therefore we
restrict ourselves to zero temperature and focus on the ground state
of the Hamiltonian (\ref{ham}). Computing the ground state now
corresponds to finding $N$ non-overlapping paths traversing the
network from top to bottom. One has to minimize the total energy of
the whole set of the paths and not of each path individually (already
the two-line problem is actually non-separable \cite{twoline}).
This task can be achieved by applying Dijkstra's
shortest path algorithm successively on a residual graph
\cite{opt-review}.

Although the lines cannot occupy the same bond of the lattice they may
touch in isolated points as exemplified in Fig. 1. This means that the
line identification based upon a bond configuration is not
unique. Since we want to calculate the roughness of lines we need to
determine the individual lines, for which we use a local rule. In 2d
the line identification is unambiguous if we simply require that the
lines cannot cross, see Fig. 1a. The same rule is applied in 3d when
the line segments ending in a single site are within the same plane
(see Fig. 1b), otherwise the choice is random (see Fig. 1c).

\begin{figure}[t]
\includegraphics[width=0.9\columnwidth]{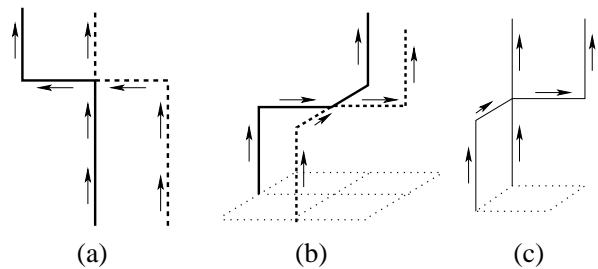} 
\caption{\label{linedef} Line identification schemes in $2$ and $3$
 dimensions. In (a) and (b) the bold and dotted lines denote the choice
 identification of two lines. (c): an example of the case where lines
 are identified randomly.}
\end{figure}

\section{Roughening in 2d}

The main quantity we are interested in is the disorder averaged line
roughness. The mean square displacement of a single line with index
$i$ in one sample is defined as
\be
w_i^2=\overline{r^2}_i-\overline{r}_i^2
\label{rough}
\ee
with $\overline{r^n}_i=H^{-1}\sum_{z=1}^H r_i^n(z)$ for $n=1,2$.  The
mean square displacement of all lines in one sample is
$\bar{w}^2=\frac{1}{N}\sum_{i=1}^N w_i^2$ and the roughness $w$ is defined
as the square root of the disorder averaged sample mean square
displacement $w=\sqrt{\langle\bar{w}^2\rangle}$.

The roughness of a one line system ($N=1$) scales as $w\sim H^{\zeta}$
in the limit of infinite transverse system size. In 2d the value of
the roughness exponent is $\zeta=2/3$ \cite{one-line-2d}, whereas in
3d it is close to $\zeta=5/8$ \cite{one-line-3d,barabasi95}. In the 
case of  a non-vanishing line density one expects to observe this
single line behavior as long as the transverse fluctuations of the
individual lines are smaller than the average line-to-line distance
$a$, which is given by the line density $\rho=1/a$ in 2d. This means
that we expect $w\sim H^\zeta$ for $H\ll a^{1/\zeta}$.

Once the transverse fluctuations of the individual lines have reached
the size of average line-to-line distance one expects a collective
behavior of the lines that restricts the individual line roughness due
to the presence of the others. If the line system behaves like an
elastic medium the roughness in the collective regime is
expected to behave like $w\sim\ln L$ \cite{superrough}. Hence, for
fixed line density $\rho$ we expect the following scaling form
\be
w \approx a\ln(L)\cdot g_{2d}(H/(a\ln L)^{1/\zeta})\;,
\label{scale2d}
\ee
where $g_{2d}(x)$ is a scaling function with the asymptotic
behavior $g_{2d}(x)\propto x^{\zeta}$ for $x \ll 1$, corresponding
to the single-line behavior, and $g_{2d}(x)=const.$ for $x \gg 1$,
corresponding to the collective regime. Note that we have assumed that
the line density enters this form only via a rescaling of the 
lateral length scales.

\begin{figure}[t]
\includegraphics[width=0.75\columnwidth,angle=-90]{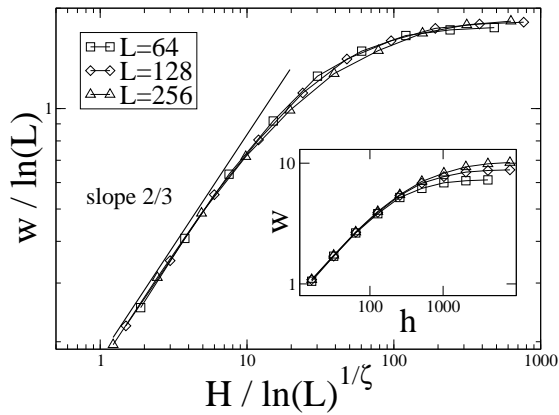}
\caption{\label{r2d_1} Scaling plot of the roughness in 2d according to the 
finite size scaling form (\ref{scale2d}). The line density is $\rho=0.05$,
the roughness exponent is $\zeta=2/3$.}
\end{figure}

\begin{figure}[t]
\includegraphics[width=0.75\columnwidth,angle=-90]{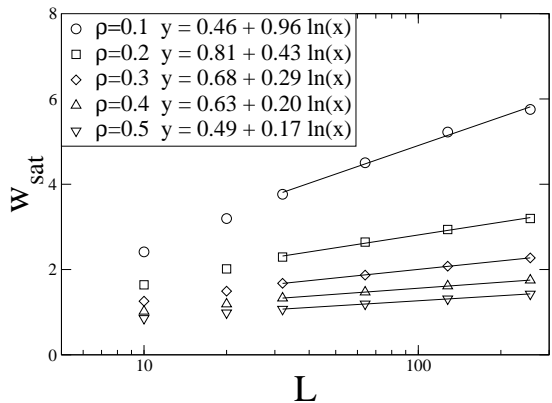}
\caption{\label{r2d_2} Saturation roughness as a
 function of the system width with fits of the form 
 $y=A+B\, {\mathrm ln}(L)$.} 
\end{figure}

\begin{figure}[t]
\includegraphics[width=0.75\columnwidth,angle=-90]{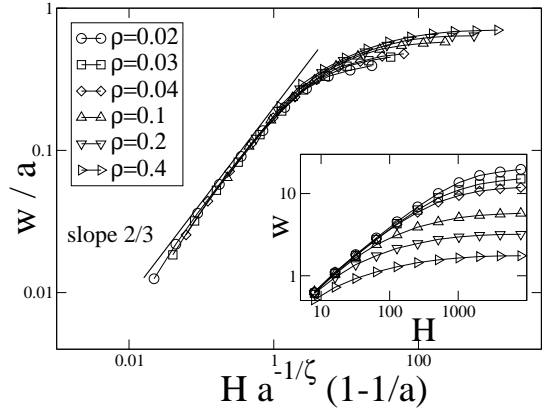}
\caption{\label{r2d_3} Scaling plot for the roughness in 2d 
according to the scaling form (\ref{scale2da}). The system size 
is fixed at $L=256$.}
%{\bf Viljo, could you replace $\rho$ by $1/a$?}}
\end{figure}

We computed the ground states of a large number of disorder
realizations (up to $10^3$) for various values of $L$, $H$ and $\rho$
and produced scaling plots for fixed line density $\rho$ according to
the suggested size scaling form (\ref{scale2d}). We found a good data
collapse for all values of $\rho$ that we checked ($\rho=0.05$,...,$0.5$). 
In Fig.\ \ref{r2d_1} we show the data collapse
for $\rho=0.05$. This particular value results in the
best data collapse for the achieveable system sizes.

%{\bf here, change the actual $\rho$, done. should one say that larger
%$\rho$ require corrections due to constant $A$ in Fig. 3 ?}

We estimated the saturation roughness $w_{\rm sat}=\lim_{H\to\infty}w(H)$
from the flat tail of the roughness curves $w(H)$ and show them in
Fig.\ \ref{r2d_2} as a function of $L$ for several values of the
line-density $\rho$. The data can be fitted to the form
$y=A+B\mathrm{ln}(L)$ again reflecting the collective super-rough
scaling of the roughness in 2d. The slope $B$ of the data sets
decreases as $\rho$ increases while the constant part $A$ does not
vary much. This leads to the strengthening of finite size effects with
increasing $\rho$.

The crossover from single-line to multi-line scaling takes place
when $H^\zeta\sim a$, where $a$ is the average line-to-line
distance $a=1/\rho$ in 2d. For fixed but large lateral system size $L$
the scaling form (\ref{scale2d}) predicts
\be
w\approx a\cdot \tilde{g}_{2d}(H/a^{1/\zeta})
\label{scale2da}
\ee
where the scaling function $\tilde{g}_{2d}(x)$ has the asymptotic
behavior $\tilde{g}_{2d}(x)\sim x^{\zeta}$ for $x \ll 1$ and
$\tilde{g}_{2d}(x) \sim const.$ for $x \gg 1$. Fig.\ \ref{r2d_3}
shows the corresponding data collapse of the roughness data that we
computed. Note that the height has been also rescaled with
a factor of $1-1/a$ to account for the limit $\rho \rightarrow 1$.

\section{Roughening in 3d}

In this section we present our numerical results and the corresponding
scaling laws obtained for the 3d case. If the observations we made in
the last section could be carried over to the 3d case, one would
expect 2 regimes: One for small heights $H$, in which the transverse
fluctuations of the lines are still much smaller than the average
line-to-line distance $a=1/\rho^{1/2}$ in 3d; and one for large $H$, in
which the line roughness is restricted due to collective effects. 
{\em Only in the case}
our line system would also for 3d fall into the same universality
class as a 3d elastic medium (this is, as we have shown the case in
2d), one would expect $w_{\rm sat}\propto\sqrt{\ln L}$ 
\cite{elastic3d,elnum3d}.

\begin{figure}[t]
\includegraphics[width=0.75\columnwidth,angle=-90]{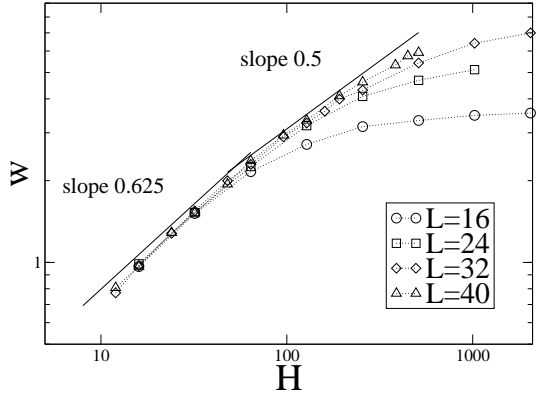}
\includegraphics[width=0.75\columnwidth,angle=-90]{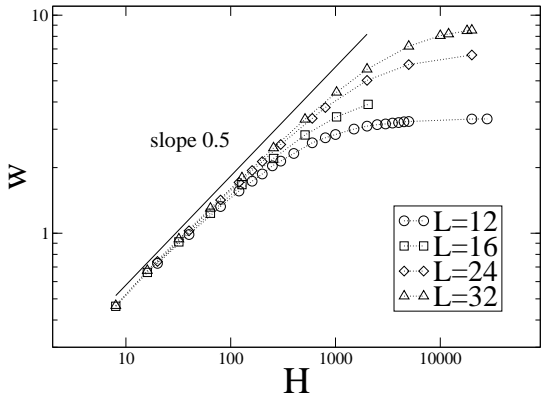}
\caption{\label{r3d_0}
Data for the roughness $w$ as a function of the height $H$ for
different transverse system sizes $L$ and two densities: $\rho=0.005$
(top) and $\rho=0.4$ (bottom). In the low density limit (top) the
crossover from single line to collective behavior is visible -
indicated by the two straight lines with slope $0.625$, the sinle line
roughness exponent, and $0.5$, respectively. In the high density limit
the crossover from collective line behavior (indicated by the straight
line with slope $0.5$) to the saturation regime is visible.
The data points are averaged over 100-1000 samples.}
\end{figure}

However, surprisingly we find i) three regimes instead of two
(c.f. the data shown in Fig. \ref{r3d_0}), and ii)
$w_{\rm sat}\propto L$, i.e. the size of the transverse fluctuations
is not restricted by the presence of a large number of other lines but
only by the lateral system size. Apparently in 3d the lines become
transparent to each other, and the wandering of any line to the
transverse direction does not induce collective behavior.

\begin{figure}[t]
\includegraphics[width=0.75\columnwidth,angle=-90]{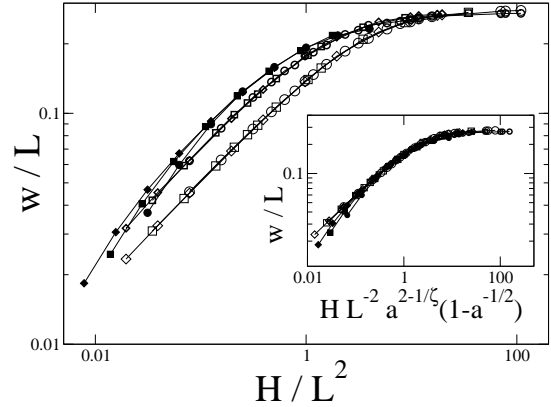}
\caption{\label{r3d_1} 
  Scaling plot of the roughness in 3d according to the finite size
  scaling form (\ref{scale3d}) for different line densities $\rho=0.1$
  (filled symbols), $0.2$ (bold) and $0.4$ (empty) for
  system sizes $L=16$ (ovals), $24$ (squares) and $32$ (diamonds). The
  inset shows data collapse according to the scaling form
  (\ref{scale3d_rho}).}
\end{figure}

\begin{figure}[t]
\includegraphics[width=0.75\columnwidth,angle=-90]{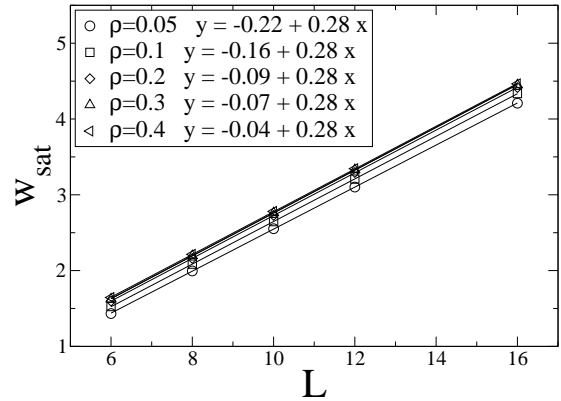}
\caption{\label{r3d_2} 
 Saturation roughness as a function of the system width with linear fits
 to the data points.} 
\end{figure}

The three regimes that we find can be characterized as follows: 1) A
single line regime for $H\ll a^{1/\zeta}$ in which the roughness
behaves as in the the one-line case: $w\propto H^\zeta$. 2) An
intermediate regime for $a^{1/\zeta}\ll H\ll L^2$ in which the
roughness increases as $H^{1/2}$, which is identical to the behavior
of random walks. 
%{\bf here, add a figure about the raw data. 
%The problem is that at L=32(too small) one can not wery well observe 
%all those different scaling regimes for a single $\rho$, see the inset 
%of Fig. \ref{r3d_3}.} 
Between the two regimes one can see a cross-over that can
be shown to be related to the entropic repulsion of the lines.
Recall that this leads in 2d asymptotically to collective 
effects, but here the consequences are different.
3) The saturation regime for $H\gg L^2$ in which
the roughness saturates at the lateral system size: $w\approx L$.
In the following we support this central result with the data 
we obtained from our ground state calculations for the 3d system
and derive the appropriate scaling forms for the different regimes.

In Fig. \ref{r3d_1} we show our results for the roughness in 3d in the
crossover region from the intermediate or multi-line regime to the
saturation regime. We show data for three different line density
values, but we have also data for other values, and they all fit well into
the scenario that we propose now). The finite size scaling plots
yield an excellent data collapse using the scaling form:
\be
w=L\cdot g_{3d}^{(a)}(H/L^2)\;.
\label{scale3d}
\ee
The scaling function $g_{3d}^{(a)}(x)$, which still depends on $a$, or
the line density $\rho=1/a^2$, has the following asymptotic behavior:
$g_{3d}^{(a)}(x)\propto x^{1/2}$ for $x\to0$ and
$g_{3d}^{(a)}(x)=const.$ for $x\to\infty$. 

The first crucial observation here -- and the essential difference to
the 2d case -- is that in the limit $L\to\infty$ the roughness is not
significantly restricted by the presence of the other lines but
approaches a value proportional to the lateral system size. Actually,
as we see from the plot of the saturation roughness as a function of
$L$ shown in Fig. \ref{r3d_2} that $w_{\rm sat}=\lim_{H\to\infty}
w(L,a) = 0.28\cdot L + c_a$, where $c_a$ is a small constant that
varies only slightly with $a$. This variation with $a$ is a boundary
effect: The free boundary conditions act effectively in a repulsive way
on the lines that competes with the steric inter-line
repulsion. Therefore systems with a lower line density show smaller
transverse line fluctuations than those with a higher density. 

The second crucial observation is that the roughness of the lines in
the intermediate regime grows like $H^{1/2}$, i.e. they have a
roughness exponent that is smaller than the single line value of
$\zeta=0.625$ and is identical to the value for simple random
walks. Although the actual line configuration is constructed in a
highly non-trivial manner via a global criterion, namely the
computation of the global $N$-line ground state, their universal
geometric properties appear to be similar to that of random walks.

\begin{figure}[t]
\includegraphics[width=0.75\columnwidth,angle=-90]{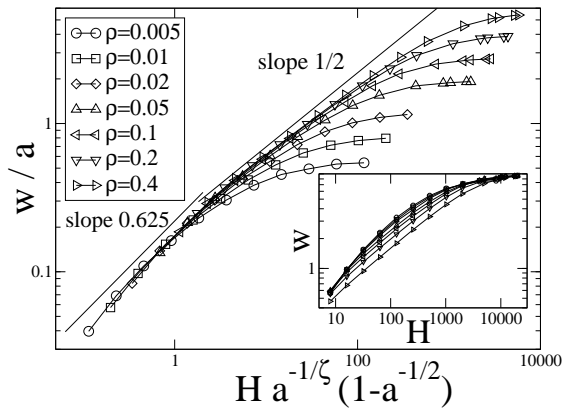}
\caption{\label{r3d_3} Scaling plot for the roughness in 3d in the 
crossover region from single to multi-line behavior. The system size 
is $L=32$. The inset shows the original, unscaled data.}
%{\bf Viljo, could you replace $\rho^{1/2}$ by $1/a$?}}
\end{figure}

The density dependence of the scaling functions $g_{3d}^{(a)}(x)$ can 
be worked out by matching it with the scaling form for the single- to
multi-line regime. Here the relevant length scale in the $H$-direction is 
$a^{1/\zeta}$, and in analogy to the 2d case we expect for $L\gg a$ 
the $L$-independent scaling form
\be
w=a\cdot\tilde{g}_{3d}(H/a^{1/\zeta})
\ee
with the asymptotics $\tilde{g}_{3d}(x)=x^\zeta$ for $x\ll 1$ and
$\tilde{g}_{3d}(x)=x^{1/2}$ for $x\gg 1$. In Fig. \ref{r3d_3} we show
the corresponding scaling plot for the data that we obtained from our
calculations. Hence we get the expected single line behavior $w\sim
H^\zeta$ when $H\ll a^{1/\zeta}$, and we obtain
\be
w=a^{1-(1/2\zeta)} H^{1/2}\quad{\rm for}\quad w\ll L
\quad{\rm and}\quad H\gg a^{1/\zeta}\;.
\label{int}
\ee
From this the natural scaling variable for the crossover region from
the intermediate regime (where $w$ should be described by (\ref{int}))
to the saturation regime (where $w$ should be proportional to $L$
according to (\ref{scale3d})) appears to be
$a^{1-(1/2\zeta)}H^{1/2}/L$ or $H/a^{1/\zeta-2}L^2$, which implies
that (\ref{scale3d}) can be rewritten as
\be
w=L\cdot g_{3d}(H/a^{1/\zeta-2}L^2) 
\quad {\rm for}\quad H\gg a^{1/\zeta}\;.
\label{scale3d_rho}
\ee
with $g_{3d}(x)=x^{1/2}$ for $x\to0$ and 
$g_{3d}(x)=0.28$ for $x\to\infty$ (see the inset in Fig. \ref{r3d_1}).
As in 2d, for high line densities  $\rho>0.1$ ($a\lesssim3$) one has to take
into account the limiting case $\rho=1$ where lines fill all parallel
lattice bonds resulting zero roughness. This limit can be incorporated
into (\ref{scale3d_rho}) by rescaling $H$ by $1/(1-\rho)$. 

\begin{figure}[]
\subfigure[]{
\includegraphics[width=0.75\columnwidth,angle=-90]{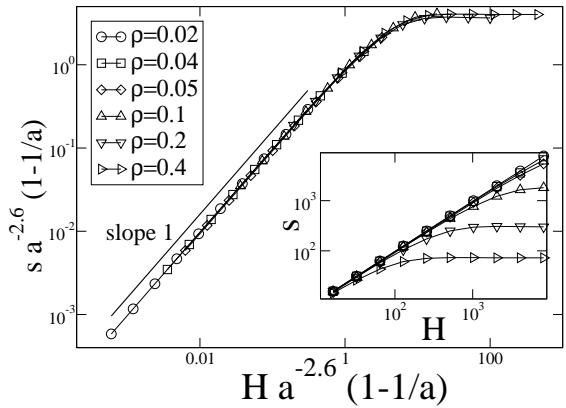}
}
\subfigure[]{
\includegraphics[width=0.75\columnwidth,angle=-90]{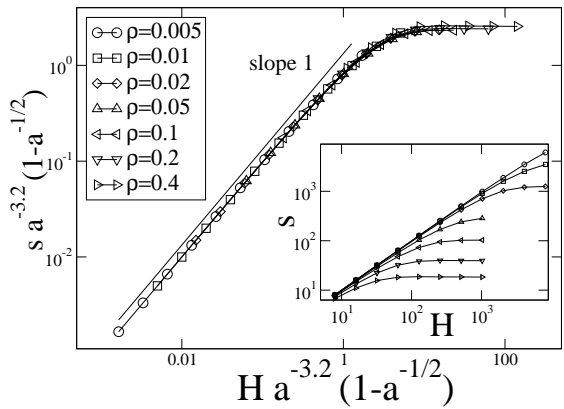}
}
\caption{\label{s2d_3d} Scaling plots for the length of line segments 
$s$ between two crossings (a) in 2d, the system size is $L=256$ and (b)
in 3d, the system size is $L=32$. The insets show the original,
unscaled data.} 

\end{figure}

At this point we would like to stress that the random walk like
scaling is {\it not} related to the actual distance at which the lines
touch or cross each other (and are in some cases continued
randomly). Both in 2d and in 3d the typical length scale $s$ between
two consecutive intersection points on one line is much larger then
the length scale $\xi$ for the crossover from single line to
collective behavior and its divergence with the line average distance
$a$ is much stronger. This can be seen in Fig. \ref{s2d_3d},
where we show scaling plots for the average length of line segments
$s$ between two crossings, from which one conludes that $s$ scales
with $a$ as $a^{2.6}$ in 2d (compared to $a^{1.5}$ for the crossover
lenght scale $\xi$) and $a^{3.2}$ in 3d (compared to $a^{1.6}$ for
$\xi$).  This also demonstrates that the lines do {\it not} behave
like independent random walkers in 3d but reflect the effect of a
steric repulsion that tends to avoid random crossings between them
(visible in 2d and in 3d).

\begin{figure}[h]
\includegraphics[width=0.3\columnwidth,angle=0]{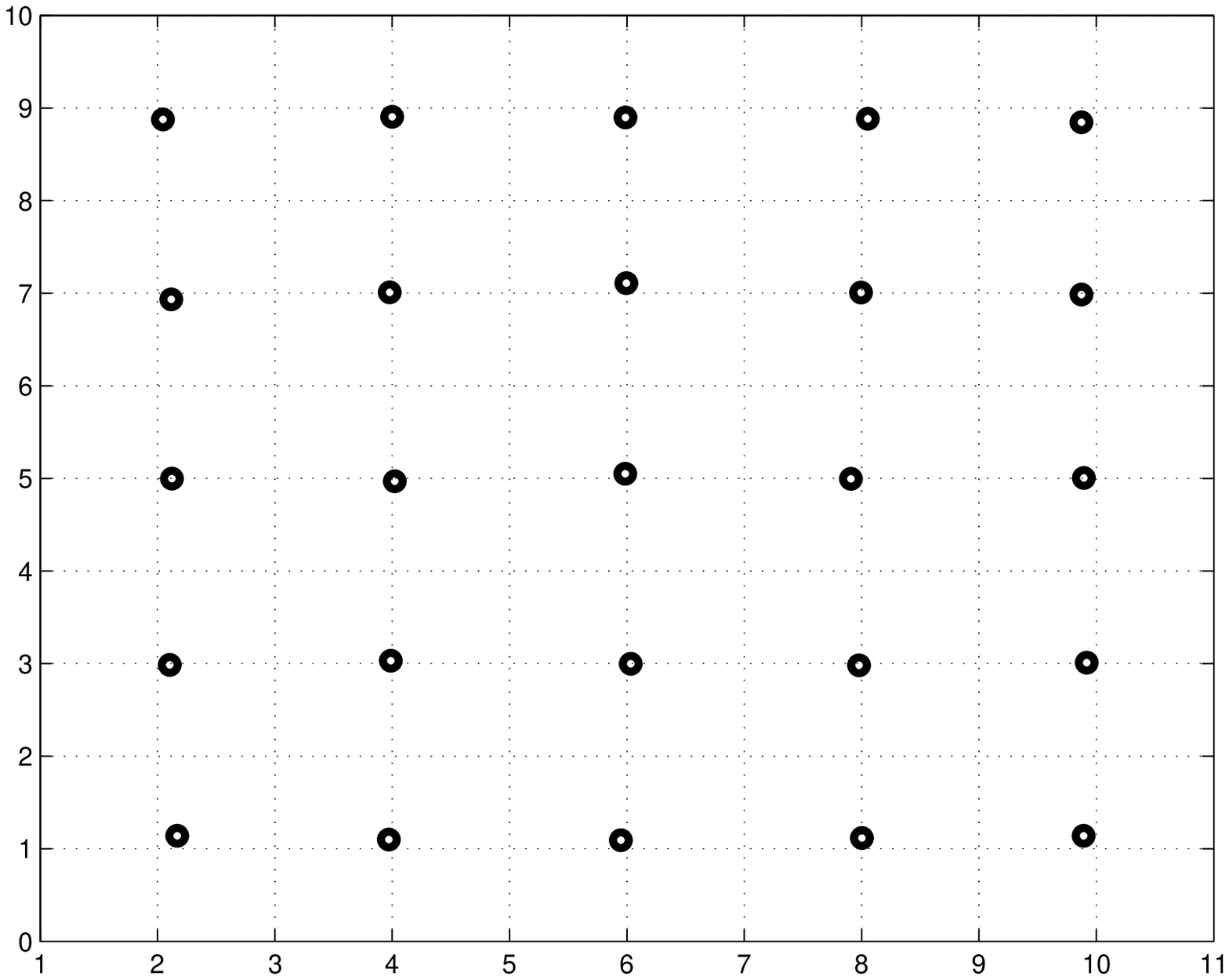}
\includegraphics[width=0.3\columnwidth,angle=0]{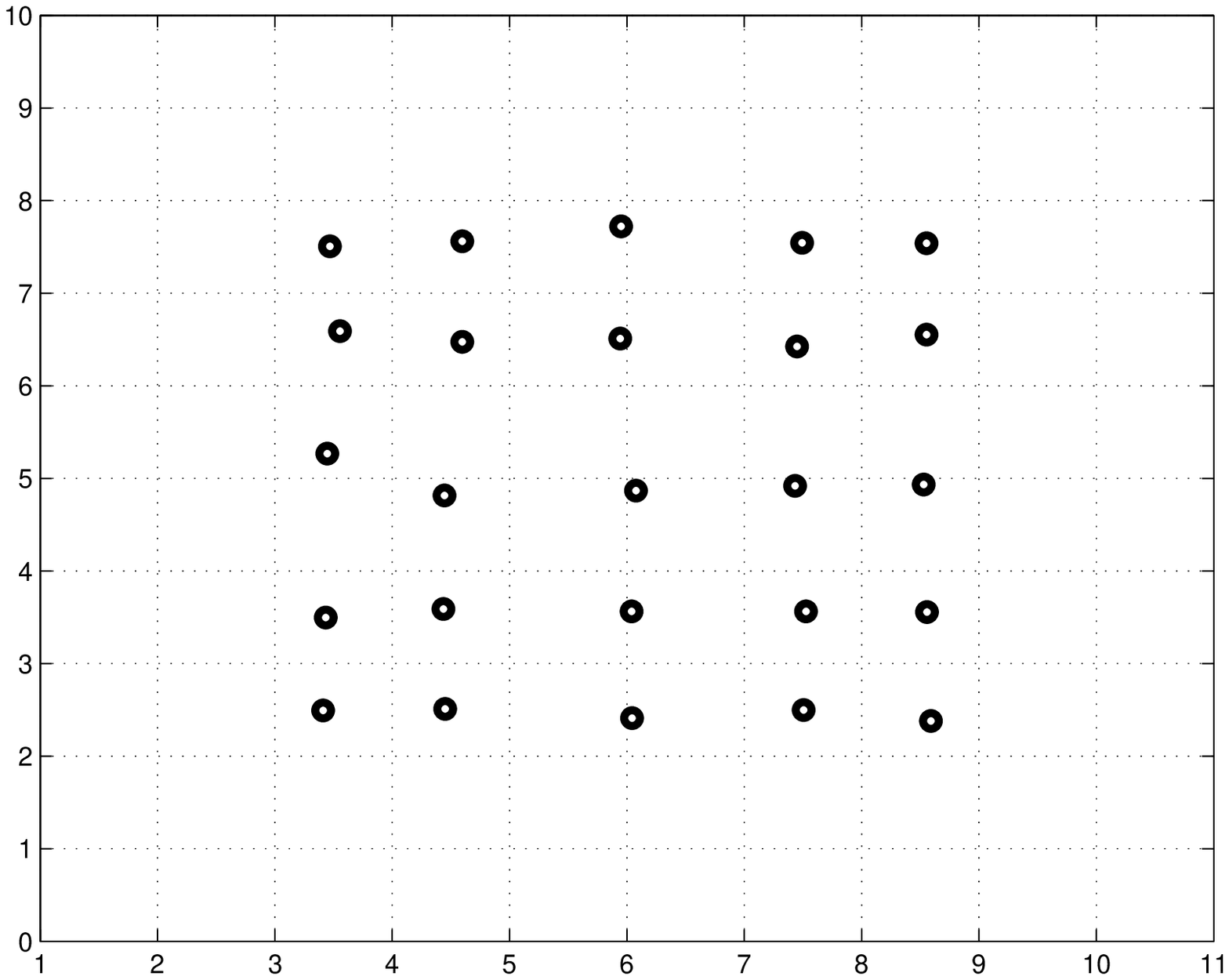}
\includegraphics[width=0.3\columnwidth,angle=0]{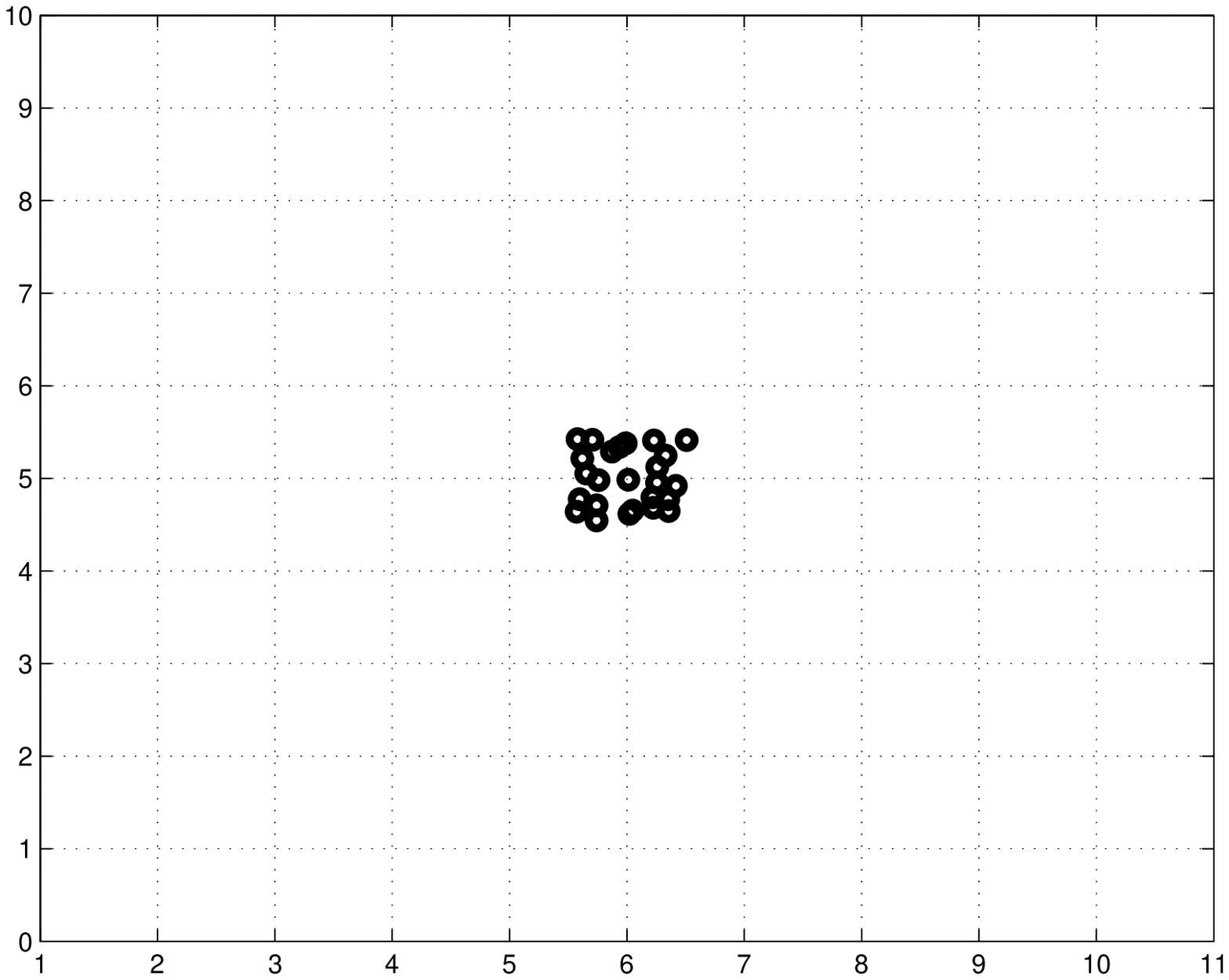}
\caption{\label{mc_ave} Average positions of center of masses of lines
 with fixed starting points. From left: H=10, H=100, H=1000.
Note that the center of mass position is averaged over the disorder,
here 1000 samples. 
}
\label{cm}
\end{figure}

The main result of calculations for the 3d system is that the lines
with only hard core repulsion can transverse the whole system, in
marked difference to the 2d case. One can visualize this result also
by looking at the disorder averaged position of the center of mass of
the individual lines (which is $\overline{r}$ as defined under
(\ref{rough})). In 2d they constitute a regular array on the base line
with lattice spacing $a$. In 3d, as we show in Fig. \ref{cm}, the
still constitute a regular array, but it concentrates, with
increasing height, more and more in the central region of the basal
plane of the system. In the limit $H\to\infty$ the average center of
mass position of each individual line will be exactly at the center of
the system since nothing restricts it from transversing the system from
one side to the other.

\section{Weak and anisotropic disorder distribution}

Next we check the robustness of the universal scaling forms in
Eqs.~(\ref{scale2d}), (\ref{scale2da}), and (\ref{scale3d}) 
against a more general disorder distribution. The random energy
$e_i$ is distributed in $[0,\delta]$ with $\delta>0$ if the bond $i$
is in the transverse direction while in $[1-\epsilon,1+\epsilon]$ with
$0\leq \epsilon\leq 1$ if the bond is in the longitudinal direction.
The disorder distributions on the transverse bonds and the
longitudinal ones are thus different unless $\epsilon=1$ and
$\delta=2$. Moreover, as $\epsilon$ becomes smaller, the energies on
the longitudinal bonds are more highly concentrated around $1$
implying weaker disorder.

\begin{figure}[b]
\includegraphics[width=8cm]{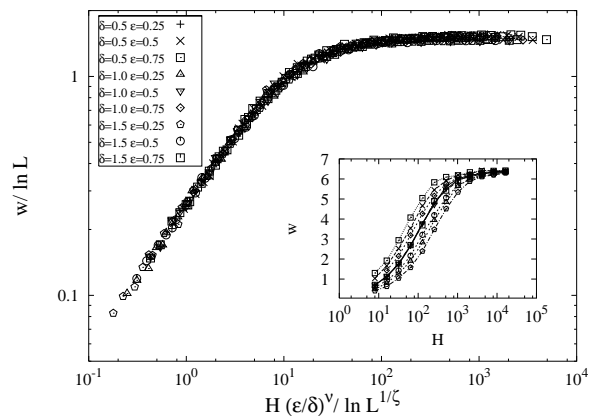}
\caption{
Data collapse of the scaled roughness $w/\ln L$ vs.  the scaling
variable $H (\epsilon/\delta)^{\nu}/(\ln L)^{1/\zeta}$ of interacting 
lines in the 2d lattice of size $L\times H$ with
$L=16$, $32$, $64$, and $128$ and the line density $N/L=1/16$.  The
parameters $\epsilon$ and $\delta$ are chosen as in the legend.  
The exact wandering exponent
$\zeta=2/3$~\cite{barabasi95} and the fitting value $\nu\simeq 0.8(1)$
are used for the scaling variable.  The inset is the plot of the
roughness $w$ vs. $H$ for $L=64$.  Dependence of $w$ on $\epsilon$ and
$\delta$ is shown before the roughness saturates.}
\label{fig:2drough}
\end{figure}

\begin{figure}[t]
\includegraphics[width=8cm]{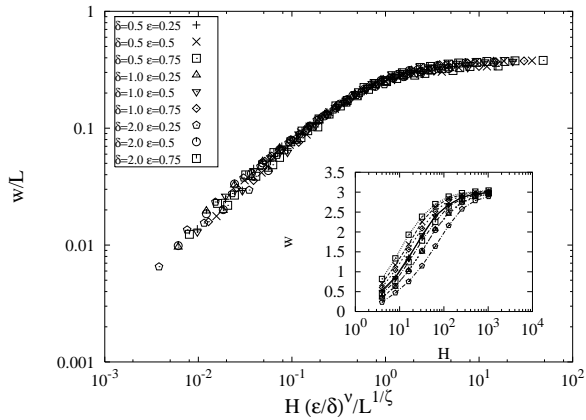}
\caption{ 
Data collapse of the scaled roughness $w/L$ vs. 
the scaling variable $H (\epsilon/\delta)^{\nu}/L^{1/\zeta}$ 
of one line in the 3d lattice of size $L\times L\times H$ 
with $L=8$, $16$, and $32$.
The wandering exponent $\zeta=0.625$~\cite{barabasi95,lassig98} is used 
and the exponent $\nu$ is found to be $0.68(8)$.
The inset is the plot of the roughness $w$ vs. $H$ for $L=8$.}
\label{fig:3drough}
\end{figure}

When $\epsilon=0$ or $\delta^{-1}=0$, the lines are flat, i.e., $w=0$
because there is no energy reduction that compensates for the energy
cost accompanying a transverse fluctuation. On the other hand, when
$\epsilon=1$ and $\delta=2$, the lines are rough and exhibits the
universal scaling behavior shown previously.  Therefore it would be
desirable to find the boundary between such distinct two limits in
the ($\epsilon,\delta$)-plane.  We computed the roughness for the
generalized disorder distribution with various values of $\epsilon$
and $\delta$, which is shown in Figs.~\ref{fig:2drough} (2d) and
\ref{fig:3drough} (3d).  We consider for 2d systems with the fixed line
density $N/L=1/16$. In 3d, a single line is studied to sort out the
effect of $\epsilon$ and $\delta$ on the roughness scaling in
Eqs.~(\ref{scale2d}), (\ref{scale2da}), and (\ref{scale3d}) more
clearly.  The roughness shows the same scaling
behaviors as in the previous sections except for the dependence on
$\epsilon$ and $\delta$ before it saturates.  It implies that 
interacting lines are rough for nonzero $\epsilon$ and $\delta^{-1}$
and that the scaling behaviors in Eqs.~(\ref{scale2d}),
(\ref{scale2da}), and (\ref{scale3d}) are universal with respect to
the variations of the disorder distribution used here.

The dependence of the roughness on $\epsilon$ and $\delta$ before it
saturates is rooted in the crossover behavior around the point
$(\epsilon,\delta^{-1})=(0,0)$.  As $(\epsilon,\delta^{-1})$ deviates
from $(0,0)$, a line roughens ($w>0$), if the reduction of
the longitudinal-bond energy obtained by the transverse fluctuation is
larger than the accompanying energy cost. By comparing the energy
reduction and cost accompanying the transverse fluctuation, one can
get the crossover (``Larkin'') length scale for $H$, $\xi_H$.
For a given line $i$, the typical fluctuation or the
standard deviation of the sum of the energies on the longitudinal
bonds, $E_{i,\parallel} = \sum_{z} e_\parallel ({\bf r}_i(z))$, is
given by $\epsilon \sqrt{H}$, which corresponds to the average
reduction of $E_{i,\parallel}$ achieved by a transverse fluctuation.
On the other hand, the line chooses the position where the cost is
minimum among $H$ possible ones ($1\leq z\leq H$) to fluctuate in the
transverse direction. The cost on the average is equal to the average
minimum value of $H$ independent random variables each of which is
distributed uniformly in $[0,\delta]$, which is given by
$\delta/(H+1)$~\cite{gumbel58}.  Comparing the energy reduction and
cost, one finds the characteristic length scale $\xi_H$ as
\begin{equation}
\xi_H \sim \left(\frac{\epsilon}{\delta}\right)^{-2/3},
\label{eq:xi}
\end{equation}
such that the roughness $w$ is zero for $H\ll \xi_H$ and
non-zero, increasing as $H^\zeta$ for $H\gg \xi_H$. From the
numerical data and the above argument, the roughness in 2d and 3d for
the generalized disorder distribution used here can be represented as 
\begin{equation}
w\sim\left\{
\begin{array}{ll}
0&(\frac{H}{\xi_H}\ll 1),\\
\left(\frac{H}{\xi_H}\right)^{\zeta} 
& (1\ll \frac{H}{\xi_H}\ll [w_{\rm sat}(L)]^{1/\zeta}),\\
w_{\rm sat}(L)& (\frac{H}{\xi_H}\gg [w_{\rm sat}(L)]^{1/\zeta}),
\end{array}
\right.
\end{equation}
with the usual roughness exponents and saturation roughness scalings
according to the dimension. The crossover scaling function
between the transient regime and the stationary state is represented
in terms of the scaling variable $H/\xi_H/(w_{\rm sat}(L))^{1/\zeta}$,
which is shown in Figs.~\ref{fig:2drough} and \ref{fig:3drough} with
the fitting value for $\nu$ in $\xi_H\sim (\epsilon/\delta)^{-\nu}$
being $0.8(1)$ (two dimension) and $0.68(8)$ (three dimension),
the latter being in rather good agreement with 2/3. The derivation of
(\ref{eq:xi}) corresponds to comparing elastic energy to the pinning
energy in a conventional derivation of the Larkin length. 
Obviously in two dimensions one should be more careful while simply
equating the costs of tranverse bonds to the elastic energy of the
line and the costs of longitudinal bonds to the strength of a pinning
potential.

\section{Conclusions}
In this work we have considered the collective effects arising
from the interaction of ``forests'' of directed polymers, in
the presence of quenched randomness. In both 2d and 3d this
is marked by a cross-over from individual, rough lines to
a ``collective'' regime with a system-size dependent roughness
$w$. In 2d, our studies augment earlier work on similar systems
using fixed line densities. The approach is based on a mapping of the problem
to the minimal matching problem of combinatorial optimization,
and the essential point of our model(s) is that the
density of lines is a free parameter. In this, we consider
the cross-over to the asymptotic collective roughness as
a function of the parallel and perpendicular disorder strengths.

In 3d, perhaps more importantly, it is found that the generally expected
logarithmic roughening is absent: while (scalar) models of 
weakly perturbed elastic manifolds do exhibit logarithmic
roughness scaling this is not the case here. An ensemble
of directed polymers with hard-core interactions exhibits
almost ``trivial'' roughening, with a random-walk -like
scaling picture. There is a crucial difference between
two and three spatial dimensions even within the confines
of the same microscopic model. An ensemble
of directed polymers with locally repulsive interactions is thus proven 
to be different from other approaches using mappings to e.g.
higher-dimensional manifolds that do exhibit logarithmic
correlations \cite{elnum3d}. It is worth noting that
this does not follow from a ``profileration of dislocations''
in a vortex lattice, but from the absence of collective
behavior.

A rather random but likewise potentially very important
consequence of the same is that in 3d 
the lines entangle - the topological state becomes highly
non-trivial such that it might be useful to describe it
in terms of knot-theoretical means \cite{entangle,Nechaev}. 
Such concomitant geometrical structures are again absent
in models for 3d elastic media. In this case, 
the description of barriers and excitations still
remains to be done including the effect of the entanglement
on both.

\acknowledgments
%Acknowledgements: 
We acknowledge gratefully the support of the European Science
foundation (ESF) SPHINX network, 
the Deutsche Forschungsgemeinschaft (DFG), and (MJA, VIP)
the Center of Excellence program of the Academy of Finland.
We thank Th. Knetter and G.\ Schr\"oder for their valuable input
during the initial stage of this work.

\end{document}